\documentclass[sigconf]{acmart}
\usepackage{subfig}
\usepackage{chngcntr}

\def\BibTeX{{\rm B\kern-.05em{\sc i\kern-.025em b}\kern-.08emT\kern-.1667em\lower.7ex\hbox{E}\kern-.125emX}}
    
\copyrightyear{2019}
\acmYear{2019}
\setcopyright{acmlicensed}
\acmConference[KDD 2019 Workshop]{AI for fashion
The fourth international workshop on fashion and KDD}{August 2019}{Anchorage, Alaska - USA}
\acmBooktitle{KDD 2019 Workshop: AI for fashion
The fourth international workshop on fashion and KDD, August 2019, Anchorage, Alaska - USA}
\acmPrice{15.00}
\acmDOI{10.1145/1122445.1122456}
\acmISBN{978-1-4503-9999-9/18/06} 

\begin{document}

%
\title{Fashion Retail: Forecasting Demand for New Items}

%
\author{Pawan Kumar Singh}
\email{pawan.ks@myntra.com}
\affiliation{%
  \institution{Myntra Designs Pvt. Ltd.}
  \city{Bangalore}
  \country{India}
}

\author{Yadunath Gupta}
\email{yadunath.gupta@myntra.com}
\affiliation{%
  \institution{Myntra Designs Pvt. Ltd.}
  \city{Bangalore}
  \country{India}
}

\author{Nilpa Jha}
\email{nilpa.jha@myntra.com}
\affiliation{%
  \institution{Myntra Designs Pvt. Ltd.}
  \city{Bangalore}
  \country{India}
}

\author{Aruna Rajan}
\email{aruna.rajan@myntra.com}
\affiliation{%
 \institution{Myntra Designs Pvt. Ltd.}
 \city{Bangalore}
 \country{India}}

%
\renewcommand{\shortauthors}{Pawan Kumar Singh, Yadunath Gupta, Nilpa Jha, and Aruna Rajan}

%
\begin{abstract}
Fashion merchandising is one of the most complicated problems in forecasting, given the transient nature of trends in colours, prints, cuts, patterns, and materials in fashion, the economies of scale achievable only in bulk production, as well as geographical variations in consumption. Retailers that serve a large customer base spend a lot of money and resources to stay prepared for meeting changing fashion demands, and incur huge losses in unsold inventory and liquidation costs \cite{H&M}. This problem has been addressed by analysts and statisticians as well as ML researchers in a conventional fashion - of building models that forecast for future demand given a particular item of fashion with historical data on its sales. To our knowledge, none of these models have generalized well to predict future demand at an abstracted level for a new design/style of fashion article. To address this problem, we present a study of large scale fashion sales data and directly infer which clothing/footwear attributes and merchandising factors drove demand for those items. We then build generalised models to forecast demand given new item attributes, and demonstrate robust performance by experimenting with different neural architectures, ML methods, and loss functions. 
\end{abstract}

%
%

%

\maketitle

\section{Introduction}
Forecasting demand for fashion retail is one of the most difficult forecasting problems in the industry, given fast changing consumer tastes, long ($>8$ months) design and production cycles, bulk manufacturing for cost efficiency, heavy competition on pricing, and increasing marketing costs. When planning for fashion merchandise, there is very little information available on what will be prevailing fashion in the future, what the competitor's mix will be, and how particular pricing and marketing interventions may need to be applied to promote merchandise. What retailers have is large volumes of previous years' sales data and they use it to forecast future purchases using conventional techniques  \cite{ellen2019}. While these help in estimating demand at reasonable levels of confidence for existing/previously sold merchandise, they cannot be used for predicting demand for new merchandise. Since multiple parameters in design interact non-linearly to define the look or appeal of an item in fashion, past sales data in itself is not instructive in predicting demand for future designs. 

In many fashion houses or retail brands, demand planning for the next season (6 months ahead) is done by  merchandisers based on their reading of the market, several visits of production and design houses, and their personal observations of what people buy. There is high variability in choices that different buyers recommend, and being limited by intuition, buyers cannot make futuristic calls on price movements and competition pressure.  Besides this, every buyer works on a narrow segment of the overall fashion merchandise (such as women's cotton kurtas), and two buyers do not interact or compare merchandise forecasts to adjust their overall forecasts. Hence, effects like product substitution, cannibalization, price-wars between different articles fulfilling the same consumer need, etc cannot be foreseen or accounted correctly.  Such inefficiencies lead to significant mismatch in the supply and demand, thus resulting in loss of business opportunity for some items, and piles of unsold inventory (working capital loss). Other than business losses, unsold inventory also leads to considerable environmental damage due to overproduction as well as disposal of unsold inventory. Hence, accurate demand forecasting well into the future of 6-8 months is crucial for better environmental health and business health. 

In this paper, we apply deep learning and tree based machine learning algorithms to get point estimates in forecasting demand for items which were not present in the catalog earlier (new or unseen items). In the next section, we briefly discuss research work related to the current problem. Section 3 explains various algorithmic variants and neural network architectures applied to the problem. Section 4 describes the data used for experiments, and the results obtained in various scenarios of modeling as well as real world deployments. 



\section{Related Work}
Traditionally, time series forecasting has been the tool-set of choice for forecasters and statisticians in retail. These models assume a continuous scenario, where historic patterns are projected into the future. For articles yet to be introduced in fashion, these methods do not hold water. 
Simpler methods of projecting new and unseen articles are discussed in \cite{ellen2019} such as average forecast, seasonal forecast, bass model, life cycle approach, etc. The bass model is an interesting diffusion based model that relies on all products having early adopters (innovators) and late ones (imitators), while the product persists for a longer duration.  Products in fashion retails are neither durable nor do they have enough life to have innovators and imitators, thus, making this model inapplicable in our scenario. We use the average forecast model as a baseline to calculate and contrast our model's performance. A comprehensive survey of demand forecast in fashion is reviewed in \cite{nenni2013demand}, however,  this does not talk about forecasting demand for new items.

New item forecast was first proposed in \cite{thomassey2006hybrid}, which uses clustering of past items' sales curve followed by assigning existing items according to a  tree based model to one of the clusters. The average sales curve of the cluster is assumed to be the sales for the new items. In our efforts to reuse this method on our fashion sales data, we found that all the items which went live on our platform at the same time grouped together, irrespective of their attributes, price and discounting. We also noticed a lack of similarity in sales behaviour of similarly clustered items even by design attributes, and visual similarity. This is intuitively justifiable, because its the combination of pricing, brand, and relative placement of a certain design which plays on a customer's mind much more heavily than either of them alone.

\section{Methodology}

While similar visual characteristics did not guarantee similar sales behaviour, our data does contain several similarly behaving time series having pricing, merchandising and visual factors in non-reducible ways. By that, we mean no intuitively explainable reduced representation for similarly behaving time series was able to be deduced. For example, conclusions like items in a particular price band and brand, or such combinatorially reducible groups behave similarly, cannot be made. The model we needed to build, thus, should learn to identify similarly behaving time series across latent parameters, and also take into account discounting, promotions, visibility variations in comparing the time series. A point in a time series is represented as
\begin{equation}
y_{it} = f(A_{i}, M_{i,t}, M_{i,t-1},..., M_{i,t-p} ,D_{i,t}, D_{i, t-1}, ..., D_{i, t-p})
\end{equation}
where $y_{it}$ is sales for item 'i' at time 't', $A_{i}$ is attribute of the item 'i' like colour - blue, material - cotton etc., $M_{it}$ indicate merchandising factors like discount, promotion for items 'i' at time 't', $D_{it}$ are derived features like trend, seasonality which are inferred from data and affect the sales, $p$ is number of time lag.

\begin{figure}[!ht] 
  \subfloat[MLP Model]{%
    \includegraphics[width=0.4\textwidth]{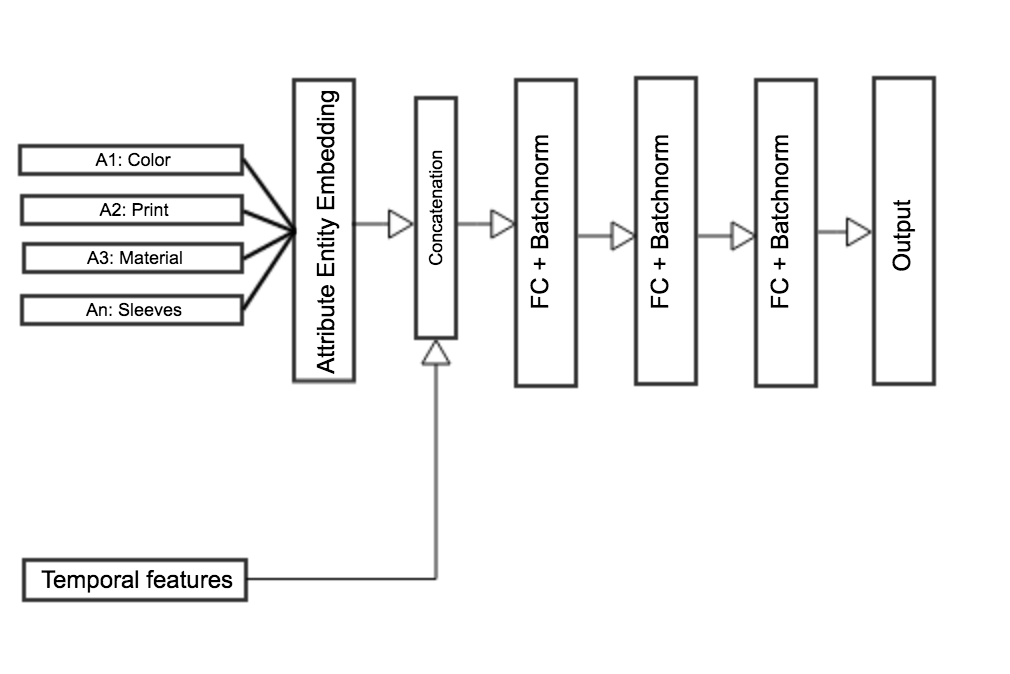} 
  } 
  \\
  \subfloat[LSTM Model ]{%
    \includegraphics[width=0.4\textwidth]{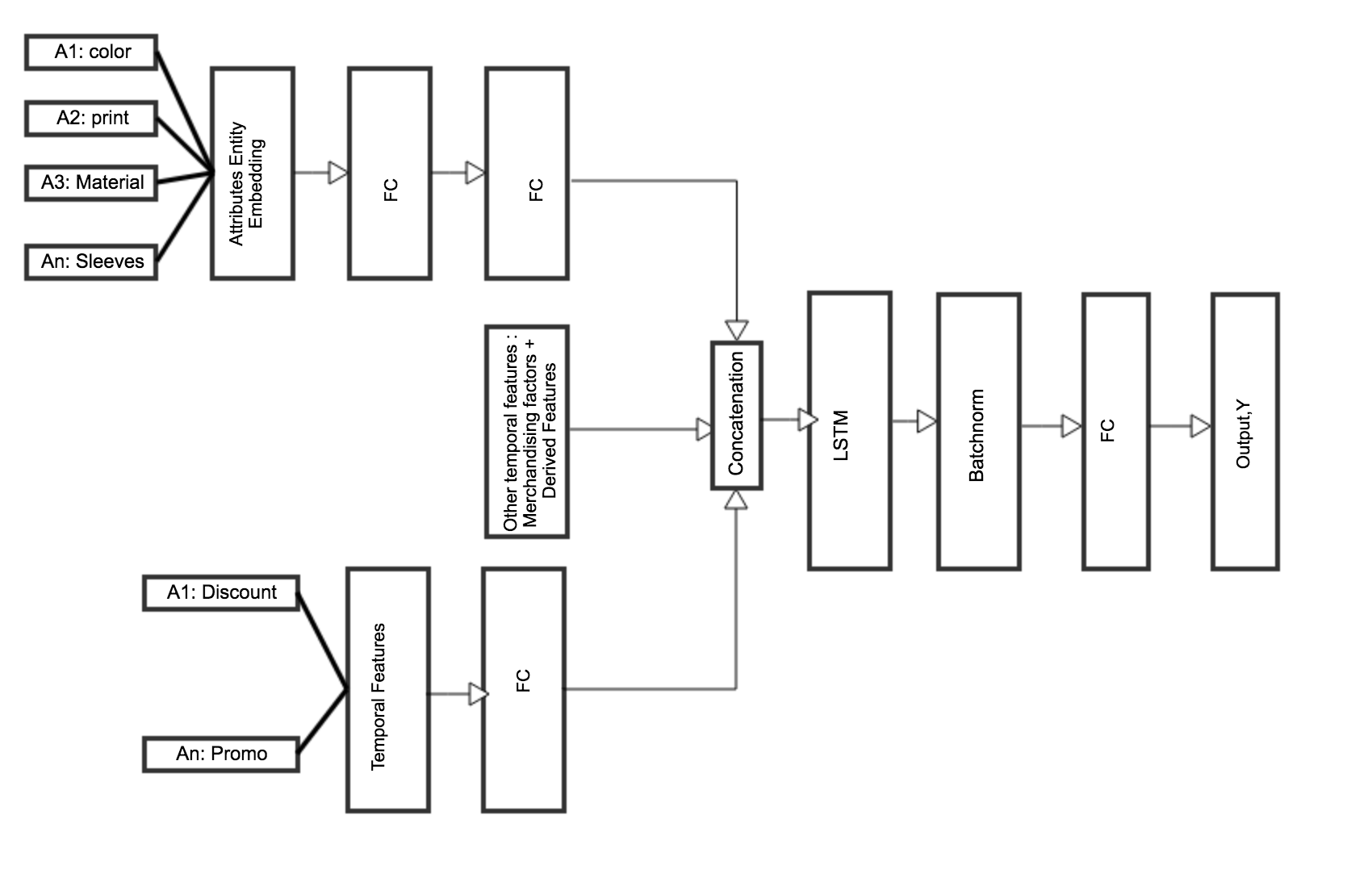} 
  } 
  \caption{DNN Model Architectures }
  \label{fig:deep_model} 
\end{figure}

As mentioned in previous section, traditional time series models are not suitable choice for $f$. Hence, we work with machine learning models ranging from tree based models like Random Forest and various flavours of Gradient Boosted Trees, to deep learning models. We train two deep learning models, first of which uses Multi Layer Perceptron (MLP) architecture, and second is based on LSTM (chosen due to its ability to model long term temporal dependencies), to derive the relation $f$. Architectures of MLP and LSTM models are shown in Fig. \ref{fig:deep_model}.

In the data, we see long tail behaviour that is typical characteristic of retail, with fewer items contributing to a majority of the sales. Due to this, we see variation of sales over several orders of magnitude. To address this high variance problem, we train our models at different scales - log and linear, and try a different set of loss functions. See Table \ref{tab:model_spec} for more details.

\begin{table*}[!ht]
  \caption{Model Specification}
  \label{tab:model_spec}
  \begin{tabular}{ccl}
    \toprule
    Model & Criterion / Loss Function \\
    \midrule
    Random Forest (RF) & Mean Squared Error (MSE)\\
    Gradient Boosted Regression Trees (GBRT) &  MSE and Huber\\
    Light Gradient Machine (LGBM) & MSE and Poisson\\
    CatBoost (CB) & MSE and Poisson \\
    XGBoost (XGB) & MSE and Poisson \\
    Attribute Embedding + MLP & MSE and Poisson\\
    Attribute Embedding + LSTM & MSE and Poisson\\
  \bottomrule
\end{tabular}
\end{table*}
Tree based and Deep learning models are chosen for their ability to model feature interactions even if transient in time, so that they capture non-linear relationship between target and regressors. Our scale is also large (\textasciitilde 1 million styles or items listed at any point in time) that limits the utility of SVM-like models that do not scale well for large sets of data and hyperparameters.  
 
Tree based models and MLP are trained in non-linear ARIMA \cite{ARIMA_wiki} manner, where lagged values of time varying features are used to capture temporal dependencies. All the data and derived features are explained in the next section. We use lagged values of temporal features up to last 4 time steps ($p=4$) . This was decided after some preliminary experiments and the intuition that temporal interactions over periods longer than 4 weeks are insignificant. Hyper-parameters of tree based models are optimized using Bayesian Hyper-parameter Optimization Technique \cite{bergstra2013making}. We use documented best practices in deep learning along with some experiments and domain understanding to choose model hyper-parameters like learning rate. A value of $10^{-3}$ was found to be effective in most cases when used with cyclic learning rates \cite{smith2017cyclical}. We have observed an improved performance of the LSTM model when Dropout \cite{hinton2012improving} and BatchNorm \cite{santurkar2018does} are used. However, to avoid over-parameterization, we do not do very extensive neural architecture search, and use a simple network, shown in figure \ref{fig:deep_model} \cite{flunkert2017deepar}. Hyper-parameter optimization is done on the validation data.

LSTM model was trained in sequence to sequence \cite{NIPS2014_5346} fashion using entire life-cycle data of a style, without explicitly coding temporal dependencies through lagged features as done with other models. We choose the LSTM approach, as several applications of this neural network architecture to sequences or time series \cite{NIPS2014_5346} have shown promising results. Our aim was to experiment with the LSTM architecture to explore how well it learns non-linear temporal patterns in the data, especially in scenarios where reduced clusters in the attribute/design space are non representative of collective behaviour. 
We create 13 models for our study, as shown in Table [1]. Performance of models are assessed on test data.

\subsection{Model Frameworks}
Deep learning models are built using deep learning framework PyTorch \cite{paszke2017automatic} , and are trained on Azure instance containing 6 CPUs and a single GPU. Well know python packages are used for Tree based models, i.e. scikit-learn \cite{scikit-learn} is used for RF and GBRT; LGBM \cite{Ke2017LightGBMAH}, CatBoost \cite{dorogush2018catboost} and XGBoost \cite{Chen:2016:XST:2939672.2939785} packages are used for other models.

\section{Experiments and Results}

\subsection{Data}
\begin{figure*}[!ht] 

  \subfloat[Sales Distribution]{%
    \includegraphics[height=0.8in,width=0.3\textwidth]{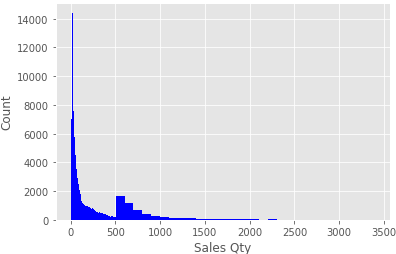} 
  } 
  \subfloat[Log Sales Distribution]{%
    \includegraphics[height=0.8in,width=0.3\textwidth]{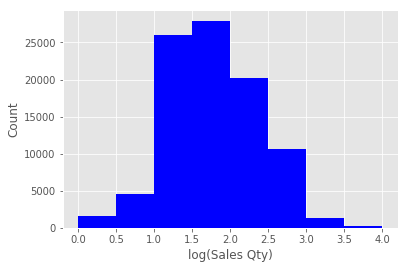} 
  } 
  \subfloat[Promotions - Sales]{%
    \includegraphics[height=0.8in,width=0.3\textwidth]{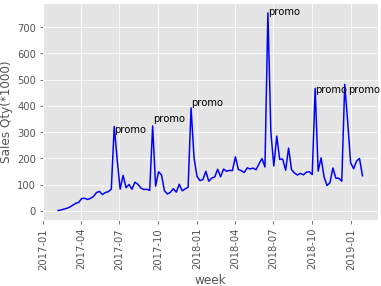} 
  } 
  \\
  \subfloat[Brand - Discount ]{%
    \includegraphics[height=0.8in,width=0.3\textwidth]{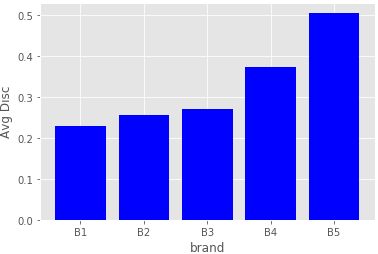} 
  } 
  \subfloat[Brand - RoS]{%
    \includegraphics[height=0.8in,width=0.3\textwidth]{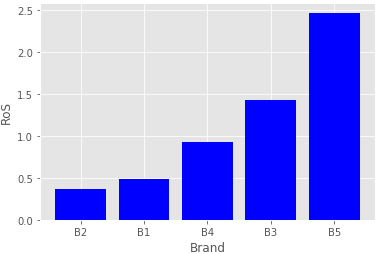} 
  }
  \subfloat[Discount - RoS]{%
    \includegraphics[height=0.8in,width=0.3\textwidth]{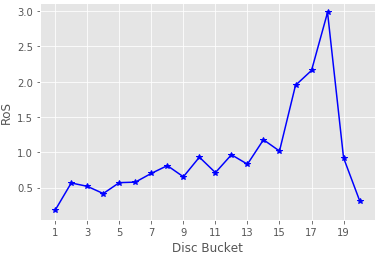} 
  }
  \\
  \subfloat[Age of Style - RoS ]{%
    \includegraphics[height=0.8in,width=0.3\textwidth]{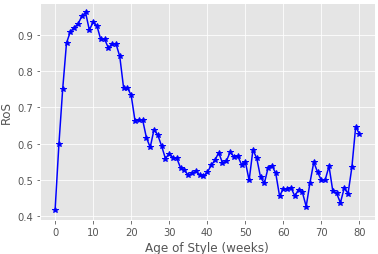} 
  } 
  \subfloat[Price Point - RoS]{%
    \includegraphics[height=0.8in,width=0.3\textwidth]{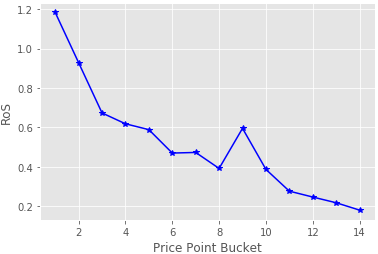} 
  }
  \subfloat[List Count Ratio - RoS]{%
    \includegraphics[height=0.8in,width=0.3\textwidth]{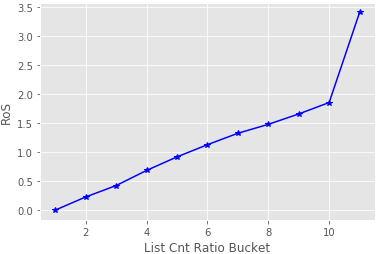} 
  } 
  \\
  \subfloat[Kurta: High RoS ]{%
    \includegraphics[height=0.8in,width=0.2\textwidth]{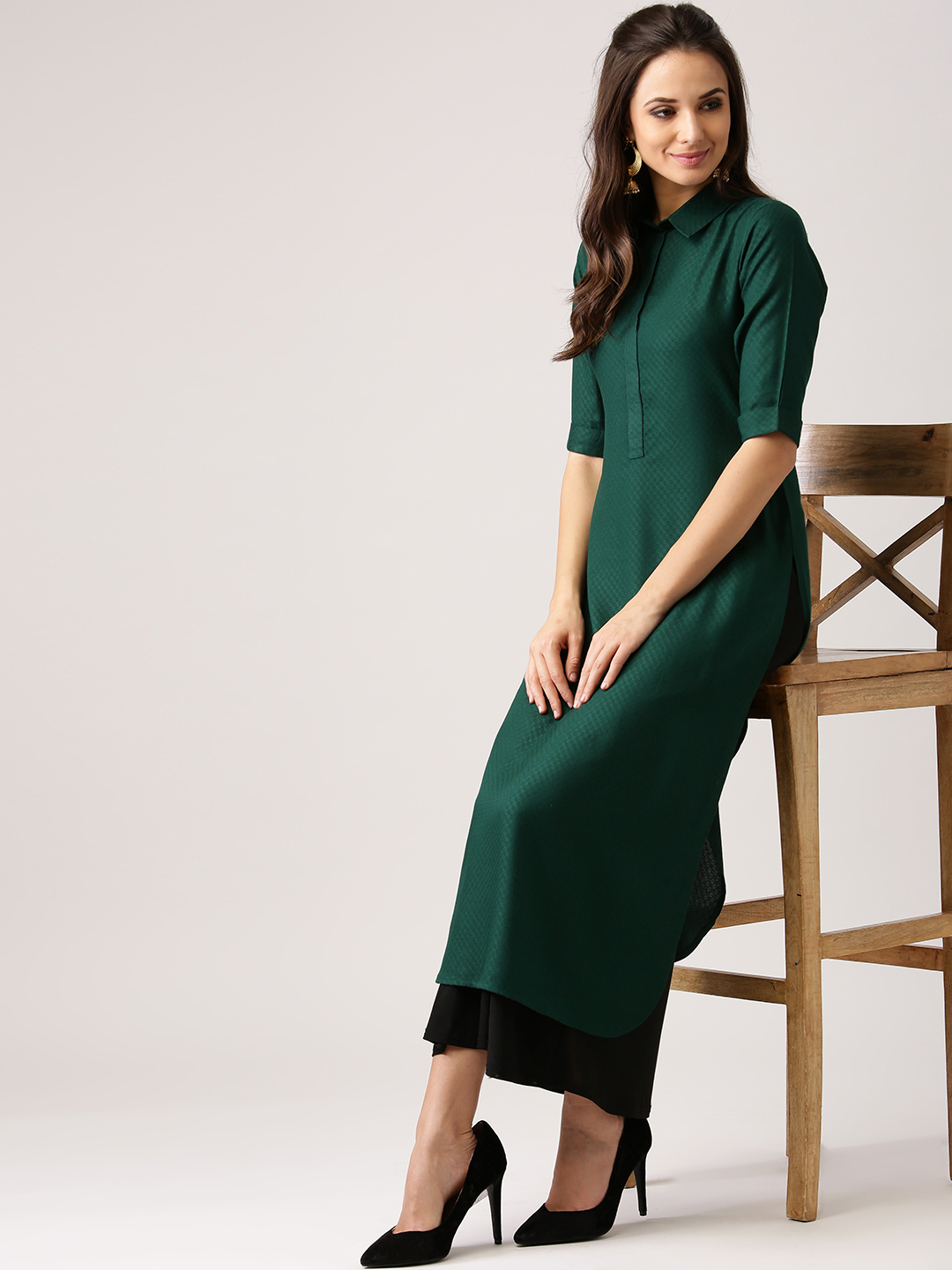} 
  } 
  \hfill 
  \subfloat[Kurta: Low RoS]{%
    \includegraphics[height=0.8in,width=0.2\textwidth]{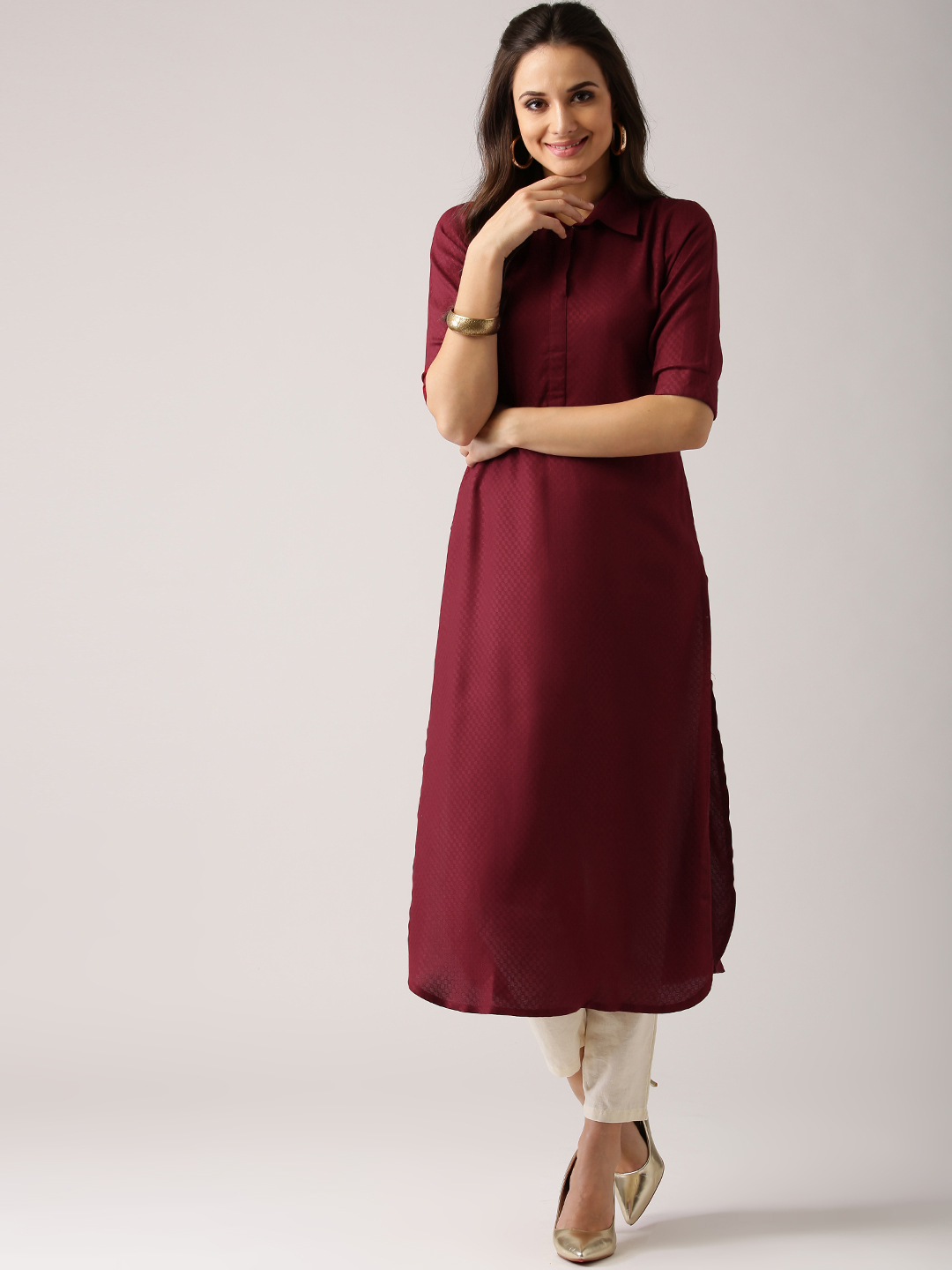} 
  }
  \hfill
  \subfloat[Shirt: High RoS]{%
    \includegraphics[height=0.8in,width=0.2\textwidth]{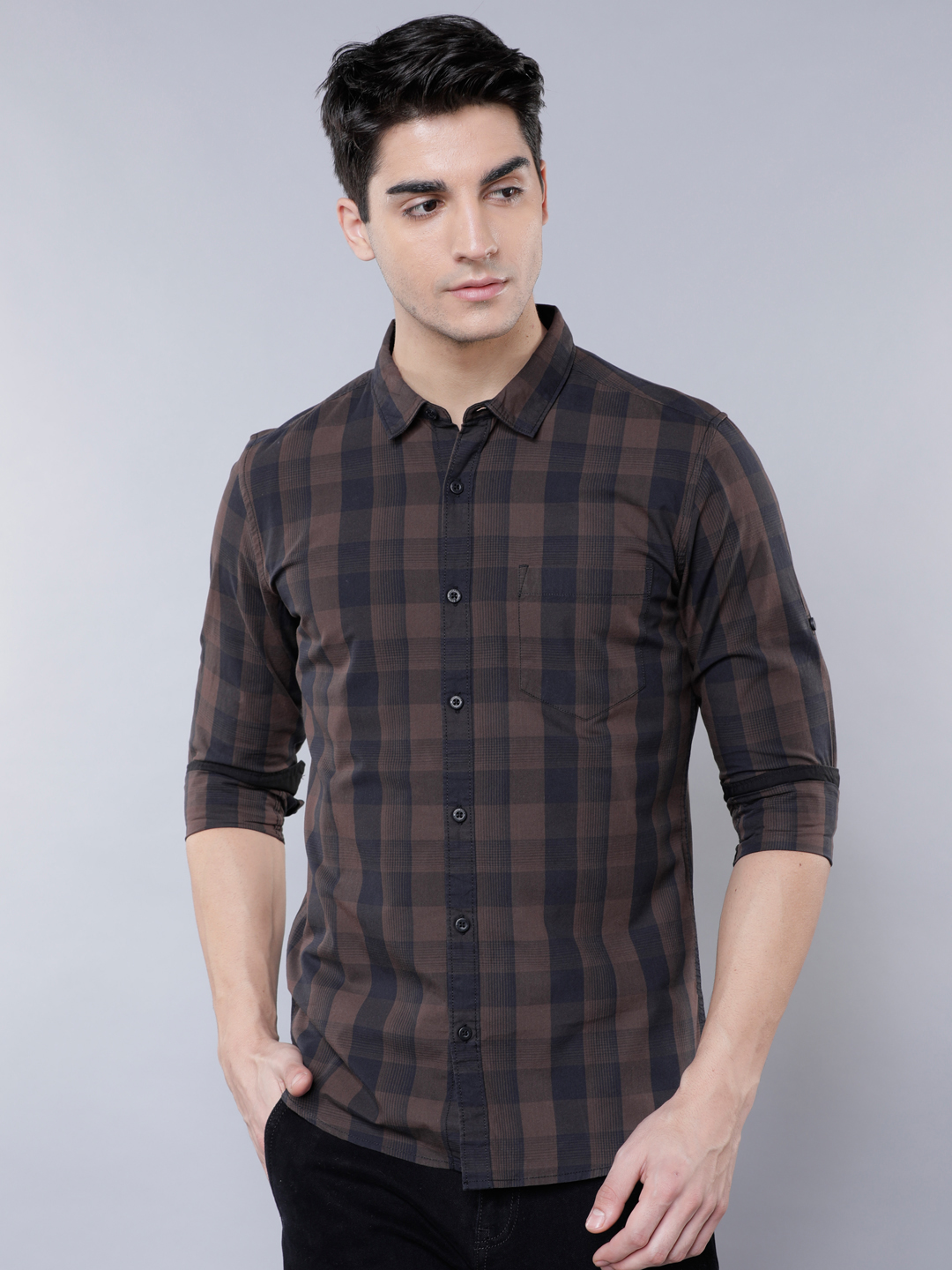} 
  } 
  \hfill
  \subfloat[Shirt: Low RoS]{%
    \includegraphics[height=0.8in,width=0.2\textwidth]{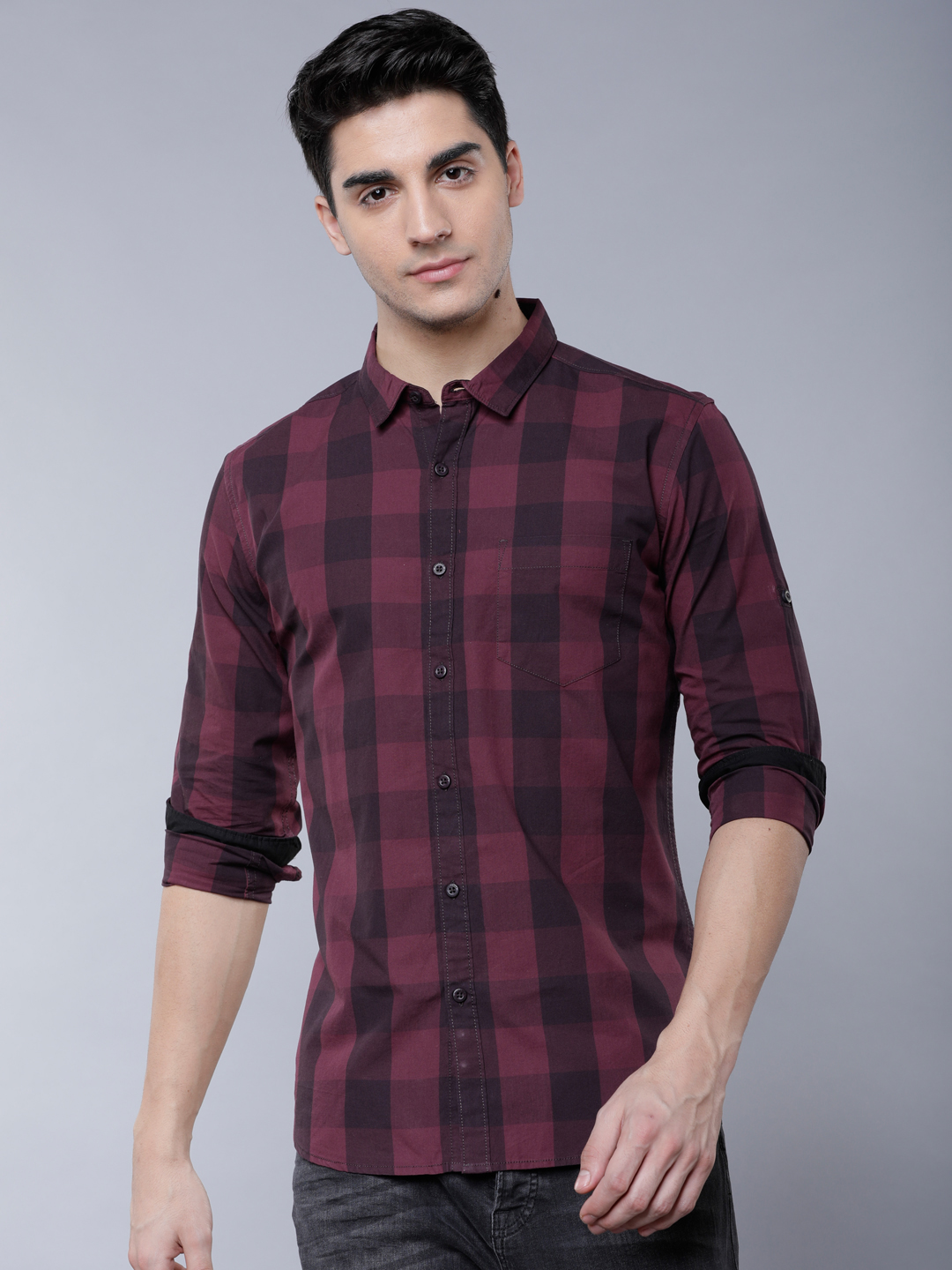} 
  } 
  
  \caption{Salient Features of Data: Sales have Poisson Distribution in linear scale (a) and Normal Distribution in log scale (b). Promotions have positive impact on sales as observed by peak in (c). Not all brands are at same discount (d), hence not all brand will have same RoS (Rate of Sales - Ratio of Total Sales and Number of days for which style was live) (e). RoS increases with discount (f), whereas it first increases and then decreases with increase in Age of Style (g). RoS is higher for lower price point (h). Higher list count ratio leads to higher RoS (i). Effect of attribute on sales can be observed by comparing (j) with (k) and (l) with (m); (j) and (k) are Kurtas form same brand, at same price point and discount, though RoS of (j) is twice that of (k), due to color difference; similarly, (l) and (m) are shirts have same brand, price point and discount, but difference in color gives (l) a RoS which is 5 times of (m)}
  \label{fig:salient_feature} 
\end{figure*}

We use historical sales data of Myntra, a leading Indian fashion e-commerce company, to train our models. Experiments are conducted on data for 5 different article types. In our fashion ontology, an article type is a hierarchy level that contains items which can be characterised by a similar set of attributes, for example - Shirts, Casual Shoes, Tops, Kurtas etc. are article types, and particular items listed under these may be referred to as style or item alternately in our work. We use data for only those items which were catalogued or went live in the last two years. Data for items which went live in the first year are taken for training, and those which went live in next 6 months are used as validation set. The validation set is used to tune hyper-parameters of the models, using standard validation techniques. Finally, a test set of subsequent 6 months was used for measuring and reporting performance. The temporal length of time series for each style will vary, as they were listed for different duration. Minimum and maximum number of time series (TS), and their minimum and maximum length available across article type are provided in Table \ref{tab:ts_details} to summarize our sequence lengths at play. Salient feature about the data and factors impacting sales are given in Figure \ref{fig:salient_feature}.

\begin{table}[!ht]
    \centering
    \caption{Time Series Details across Data}
  \label{tab:ts_details}
\begin{tabular}{p{1cm}p{1cm}p{1cm}p{1cm}p{1cm}}
 \hline
 Data & Min No. of TS & Max No. of TS & Min TS Length & Max TS Length\\
 \hline
 Train   & 12,541    &42,206 &   4 & 104\\
 Valid   & 7,489    &29,669&   4 & 52\\
 Test   & 6,732    & 18,364&   4 & 26\\
 \hline
\end{tabular}
\end{table}

We model promotions, discount, and list page views (visibility) along with fashion attributes of the style as external regressors. Some of these features are not known for future time steps at the time of prediction. Therefore we transform most of these features so that default values of promotions and discounts for future time steps can be easily approximated without remembering the training data. The details of engineered features are mentioned herein.

\begin{itemize}
 \item \textbf{Fashion Factors:}
\begin{itemize}
   
    \item Fashion related \textit{Attributes} such as colour, material etc. of a style are used. These attributes may be different for different article types. We embed \cite{guo2016entity} these attributes in order to both compress their representations while preserving salient features, as well as capture mutual similarities and differences. We learn these embeddings in the training phase. In our tree based approach, we use a simple one hot embedding  \cite{one_hot} of attributes. Attribute values with frequency less than  1\% are grouped into a dummy value to indicate values that may not be well represented in the data, as well as new and unseen values in future. 
\end{itemize}
\item \textbf{Merchandising Factors:} 
\begin{itemize}
    
\item \textit{Discount}: In our initial analysis of the data, we found that most brands sold at an average (consistent) discount on our platform, while there were intra-brand variations in discounts that sometimes boosted sales on the retail platform. We capture the discount deviation from both the brand average, and overall retail platform average, hence, as we found this feature to contain more information than the item/style's absolute discount. A value of 0 in this case for future will mean that style will be sold at average brand/ platform discount. This feature also captured the non-linear and brand specific effects of discounting in fashion retail. 
\item \textit{Visibility}: Visibility features are derived from the list page views, which is the shelf space allocated to a style in an online store. List views ratio with respect to brand and platform are numerical measures of style visibility dispersion, and have big impact on observed sales, so we use them as features. List views given to a style depend on its sales, CTR, applied promotions etc. But in the absence of this information, usually platform average list views are given to new styles. Hence a value of 1 for future time steps is a reasonable assumption except for pre-decided special promotion days where the visibility can be appropriately boosted by a factor.
\item \textit{Promotion}: To model the effect of sales drop just before and after a promotion, features like days to promotion and days from promotion are used. In Myntra and in the Indian retail scenario in general, certain country-wide observed holidays/occasions are promotional shopping festival days, such as Diwali, Valentine's day, etc. In the run up to a shopping festival (promotional), customers tend to postpone their buying till the promotional event, and immediately after a period of intense activity, we see a significant lull in shopping enthusiasm. Hence the choice of maintaining a calendar like feature to indicate a count down to and from planned promotional events. 
\end{itemize}

\item \textbf{Derived Features:}
\begin{itemize}
\item \textit{Age of Style}: Shelf life of a style. With longer shelf life, the style's demand may decay with time.
\item \textit{Trend and Seasonality}: To model a trend in interest over time, the number of weeks between experiment start date and the current date is used. In order to model seasonality in purchase patterns, first three terms of the Fourier transform of week of year are used as features. For a new item, these can be derived during prediction. 
\item \textit{Cannibalisation}: Cannibalisation is a commerce specific scenario where given that buyers/customers have a certain need, equivalent items may cannibalise each other's sales to meet that need. We create features like number of styles listed in a week, number of styles listed within the same brand in that week, number of styles listed by other brands in similar price ranges, etc. If all styles to be considered are available, along with their merchandising factors, these features can be inferred for new items; if not available  then averages/medians may be used as representative values. 
\end{itemize}

\end{itemize}
\subsection{Testing and Evaluation}
We use weighted mean absolute percentage error (wMAPE), equation \ref{eq:wMAPE}, where the weight is the actual sales realised for an item.
\begin{equation}
wMAPE = \frac{\sum_{i = 1}^{i = n}\sum_{t=1}^{t = t_{i}}|{\hat{y}_{it} - y_{it}}|}{\sum_{i = 1}^{i = n}\sum_{t=1}^{t = t_{i}}y_{it}}
\label{eq:wMAPE}
\end{equation}

$y_{it}$ and $\hat{y}_{it}$ is actual and forecasted sales of an item `i' at time `t'. $n$ is total number of items, $t_{i}$ is the length of time series for item `i'.

We choose to weight our MAPE by the item's actual sales in accordance with our tolerance for error in predicted values, so that the tolerance is lower with higher sales volumes. To illustrate the robustness of choice in wMAPE over MAPE, if actual sales for a set of items are 0, 5, and 10; and forecasted values are 1, 10 and 10; MAPE would be infinite, whereas wMAPE would be 0.4. In under-forecasting scenarios, errors are upper bounded by a wMAPE of 1; in overforecasting scenarios, wMAPEs may be arbitrarily high. We do not symmetrise our under-forecasting and over-forecasting scenarios, because over-forecasting leads to huge build up of inventory (due to need to order in lots or minimum order quantities). Generally speaking, the cost incurred per unit over-forecasted is much higher than the potential revenues missed per unit by under-forecasting. This is peculiar to retail supply chains where procurement lags are long (such as in fashion) and larger minimum order quantities apply. 

When working with fashion buyers (known as planners) to operationalize our plans and evaluate our forecasts on real buys, we learnt that a relative priority of items is important to the procurement process since procurement happens in lots of minimum order quantity. An item with low forecasted sales may therefore not be ordered due to restrictions in buying budgets, time, and inventory holding capacity. Therefore, for an item that has higher actual sales realised relative to another, the forecasted sales should also be relatively higher so that ordering it ensures higher sell through rates as well as lesser inventory pile up. To capture this, we use the Pearson correlation, equation \ref{eq:pearson} and the Kendall tau, equation \ref{eq:tau}. 

\begin{equation}
    \rho_{y_{i}, \hat{y}_{i}} = \frac{E[y_{i}\hat{y}_{i}] - E[y_{i}]E[\hat{y}_{i}]}{\sqrt{E[y_{i}^{2}] - E[y_{i}]^{2}} \sqrt{E[\hat{y}_{i}^{2}] - E[\hat{y}_{i}]^{2}}}
    \label{eq:pearson}
\end{equation}

\begin{equation}
    tau = \frac{(P - Q)}{ \sqrt{((P + Q + T) * (P + Q + U))}}
    \label{eq:tau}
\end{equation}

$y_{i}$ and $\hat{y}_{i}$ are total actual and forecasted sales of item i. P is the number of concordant pairs, Q the number of discordant pairs, T the number of ties only in $y_{i}$, and U the number of ties only in $\hat{y}_{i}$.

Pearson Correlation ensures that forecasted values and actual values move together in the same direction, and Kendall Tau takes into account relative ordering of the quantities between forecasted and actual values. 

For model tuning, we use Mean Squared Error (MSE) - equation \ref{eq:mse}, Poisson Loss - equation \ref{eq:poisson} and Huber Loss - equation \ref{eq:huber}. 
\begin{equation}
    MSE = \frac{\sum_{i = 1}^{i = n}\sum_{t=1}^{t = t_{i}}{(\hat{y}_{it} - y_{it})}^{2}}{\sum_{i = 1}^{i = n}t_i}
    \label{eq:mse}
\end{equation}
\begin{equation}
    Poisson\hspace{1mm} Loss = \sum_{i = 1}^{i = n}\sum_{t=1}^{t = t_{i}}{\hat{y}_{it} - y_{it}*log(\hat{y}_{it})}
    \label{eq:poisson}
\end{equation}

\begin{equation}
    Huber \hspace{1mm} Loss = \frac{1}{2}\sum_{i = 1}^{i = n}\sum_{t=1}^{t = t_{i}}\begin{cases}
    {(\hat{y}_{it} - y_{it})}^{2} & \text{if}\hspace{1mm} |\hat{y}_{it} - y_{it}| \leq \delta \\
    \delta* |\hat{y}_{it} - y_{it}| - \frac{1}{2}\delta^{2} & \text{otherwise}
    \end{cases}
    \label{eq:huber}
\end{equation}

In a perfect world we would have preferred to optimize on one or all of the metric- wMAPE, PerasonR, or Kendall Tau, which we use to evaluate model, for model training. But none of these metric can be arrived from likelihood function, as is the case with MSE and Poisson. Under Gaussian assumption of target variable, likelihood function and MSE gives same solution, hence MSE is most preferred loss function for problem at hand. As, evident from figure \ref{fig:salient_feature}(b) log transformed sales have Gaussian Distribution, hence MSE loss in log scale are used for model training. However, typically retail data such as ours \ref{fig:salient_feature}(a) shows long tailed distribution in linear scale, hence we use Poisson loss in linear scale for learning model parameters. Huber Loss is used to minimize the effect of outlier on the training process. For each model, we specify the loss function before tabulating the wMAPE and ranking loss values.

\subsection{Results}
Tables \ref{tab:perf_shirts_test} and \ref{tab:perf_casual_shoes_test} show performance of top five models along with naive model, for two types of articles, namely shirts and casual shoes. For completeness, performance on other article types along with performance on training data is tabulated in Tables \ref{tab:perf_kurtas_test} to \ref{tab:perf_tshirts_train} of the Appendix. We observe that almost all ML based models outperform the naive average based projection model. XGBoost with MSE loss, when optimized in logarithmic scale gives best performance followed by GBRT. Among deep learning models, LSTM with Poisson loss, when optimized in linear scale gives best performance, MLP does not feature in top 5 performers, hence metric for it is not provided.

\begin{figure*}[!ht] 

  \subfloat[Good Forecast]{%
    \includegraphics[height=0.8in,width=0.3\textwidth]{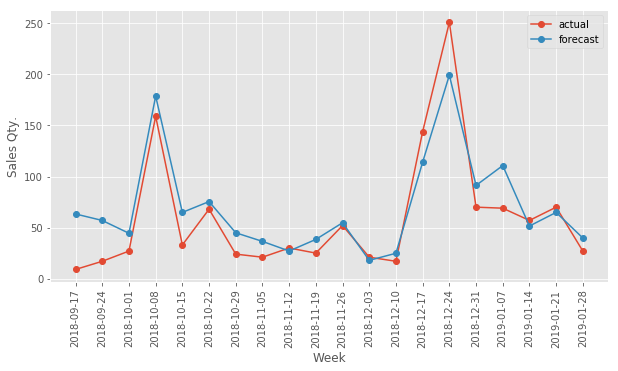} 
  } 
  \subfloat[Under Forecast for some weeks]{%
    \includegraphics[height=0.8in,width=0.3\textwidth]{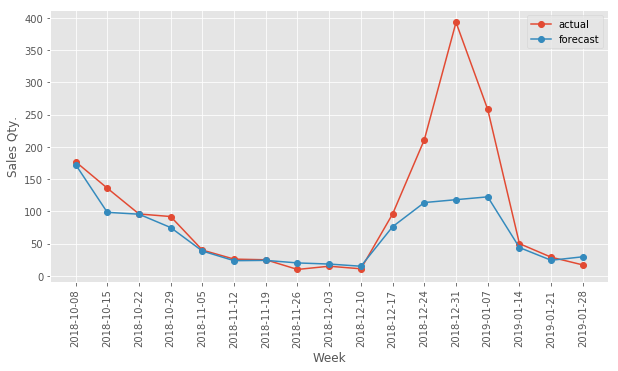} 
  } 
  \subfloat[Bad Forecast]{%
    \includegraphics[height=0.8in,width=0.3\textwidth]{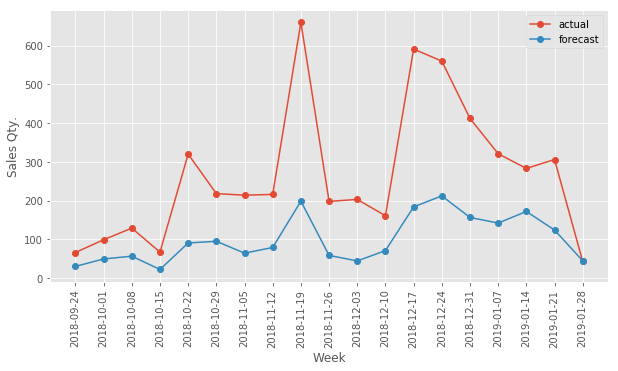} 
  } 
 
  \caption{Actual vs. Forecasted: (a) wMAPE=0.34, is an example of good forecast, (b) wMAPE = 0.37, is an example of good forecast where we are under-forecasting for few weeks, (c) wMAPE = 0.63 is an example of bad forecast or heavy under-forecast}
  \label{fig:forecast_actual} 
\end{figure*}

We provide example of good forecast - Fig. \ref{fig:forecast_actual}(a); Fig. \ref{fig:forecast_actual}(b) is an example where forecast is good for all but 3 weeks during which we under-forecast. This is explainable as sales of this style just peaks after promotion period, whereas our model learns to forecast lower just after promotion, as general trend is; Fig. \ref{fig:forecast_actual}(c) is an example of bad forecast, we are heavily under-forecasting, this is being observed because this style is an exception in terms of sales for all styles belonging to its brand. These examples tells us that even though we have used tree based and deep learning models, results of which are considered to be not easily explainable - however, using derived features we can easily explain the results of our model.

\begin{figure}[!ht] 

  \subfloat[Discount - Forecast]{%
    \includegraphics[width=0.4\textwidth]{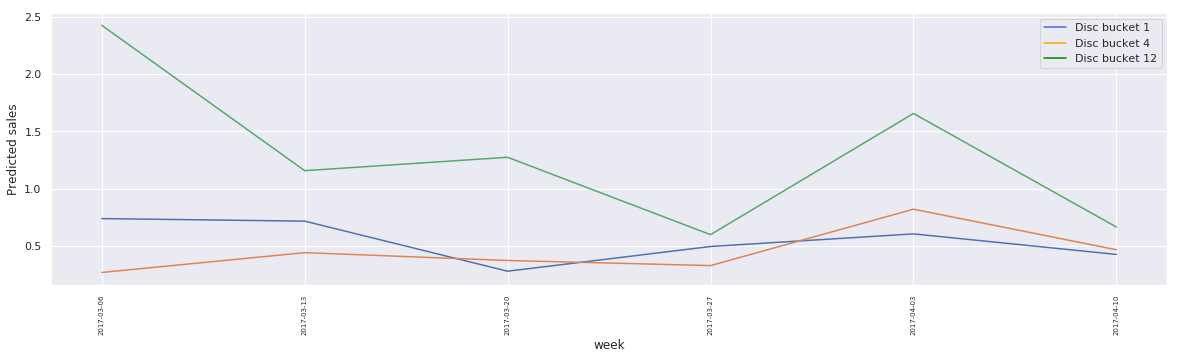} 
  } 

  \subfloat[List Count Ratio - Forecast]{%
    \includegraphics[width=0.4\textwidth]{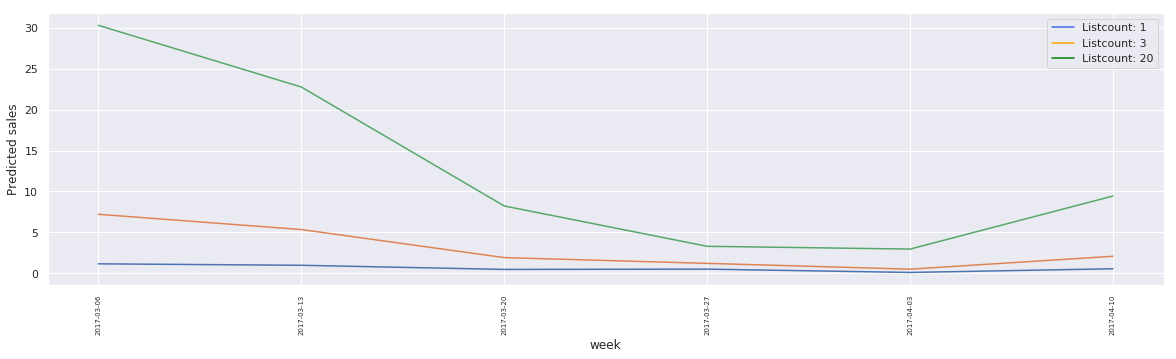} 
  } 
  
  \subfloat[Brand Style Count - Forecast]{%
    \includegraphics[width=0.4\textwidth]{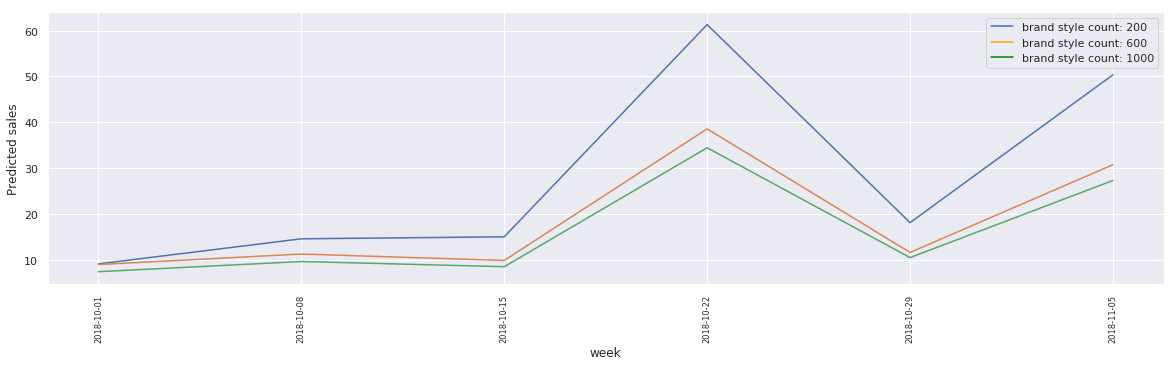} 
  } 

  \caption{Effect of derived features on forecast }
  \label{fig:sensitivity_analysis} 
\end{figure}

To illustrate the usefulness of transformed and derived features, we show forecast increases by increase in discount Fig. \ref{fig:sensitivity_analysis}(a), higher discount bucket implies higher discount. Fig. \ref{fig:sensitivity_analysis}(b) illustrates impact of increasing list count ratio on the forecast. As expected forecast increases with increase in list count ratio. Effect of cannibalization feature - brand style count, is shown in Fig. \ref{fig:sensitivity_analysis}(c), increase in number of style from a brand decreases the forecasted sales, as would be expected.

\begin{table*}[!ht]
    \centering
    \caption{Model Performance for Shirts on Test Data, 16,409 time series or items}
  \label{tab:perf_shirts_test}
    \begin{tabular}{p{3cm}| p{2cm}| p{2cm}| p{2cm}| p{2cm}| p{2cm}| p{2cm}}
    \hline
    Model & Criterion/Loss Function& \multicolumn{3}{c|}{wMAPE}& PearsonR & Kendall Tau \\
    \hline
        & &item-week & item & article type & &\\ 
    \hline
      Naive (Avg. style-week sales) Model & &0.97 & 0.82 & 0.39&0.26 &0.43 \\
XGBoost&MSE&\textbf{0.52}&\textbf{0.38}&\textbf{0}&\textbf{0.86}&\textbf{0.76}\\
GBRT&Huber&0.54&0.41&0.06&0.84&0.76\\
LSTM&Poisson&0.56&0.42&0.17&0.85&0.67\\
CatBoost&MSE&0.56&0.42&0.04&0.81&0.73\\
LGBM&MSE&0.57&0.43&0.03&0.81&0.74\\
        
    \hline
    \end{tabular}
    
\end{table*}

\begin{table*}[!ht]
    \centering
    \caption{Model Performance for Casual Shoes on Test Data, 6,732 time series or items}
  \label{tab:perf_casual_shoes_test}
    \begin{tabular}{p{3cm}| p{2cm}| p{2cm}| p{2cm}| p{2cm}| p{2cm}| p{2cm}}
    \hline
          Model & Criterion/Loss Function& \multicolumn{3}{c|}{wMAPE}&  PearsonR & Kendall Tau  \\
    \hline
        & &item-week & item & article type & &\\ 
    \hline
        Naive (Avg. style-week sales) Model & &1 &0.85 &0.43 &0.36 &0.46 \\
XGBoost&MSE&\textbf{0.51}&\textbf{0.38}&0.11&0.89&\textbf{0.74}\\
GBRT&Huber&0.52&\textbf{0.38}&0.15&0.89&\textbf{0.74}\\
CatBoost&MSE&0.52&\textbf{0.38}&0.1&0.89&0.71\\
LGBM&MSE&0.54&0.39&0.05&0.88&0.71\\
XGBoost&Poisson&0.56&0.4&\textbf{0.04}&\textbf{0.91}&0.67\\
        
    \hline
    \end{tabular}    
\end{table*}

\subsection{Deployment in industrial setting} \label{deploy}
We have tested and deployed our models for the following fashion retail use cases at Myntra-Jabong. We also talk about futuristic scenarios where we are working to deploy our models. 
\begin{itemize}
\item \textit{Seasonal assortment Planning}: Fashion Retailers have to plan their assortment a year in advance due to manufacturing lead times. At the time planners do not have any information about the actual products, so they create all plans at attribute combination level and use an average based projection together with intuitive calls to allocate inventory budget. Our model when used with appropriate simulations can generate forecasts for all possible attribute combinations of styles. This result was used to decide the set of attribute combinations on which buying budgets should be spent, for two major footwear brands during the buying season of Autumn Winter (AW) 18. We saw a year-on-year improvement of  10\% and 7\% in the overall one month sell through rate of these brands for footwear category. Sell through rate is defined as the percentage of inventory at the season start which was sold during a specified period.
\item \textit{Product Selection in Roadshows} : Wherever a catalogue of items (with their descriptions and brands) is made available for a buyer to consume in events like Fashion Roadshows which buyers frequent to find out actually available assortment from different brands, our model can quickly compute projected sales for different items present in the roadshow. A buyer may use his/her intuition in addition to our model output to get directional information on which products to spend budget on. This deployment is a work in progress at our current organisation. 
\item \textit{Drop Planning}: Purchase orders to manufacturers are placed long before the start of the season. However, to optimally utilize warehouse/ store space, deliveries / drop are taken in phased manner. Currently, all retailers plan drops at a fixed interval irrespective of how demand for an item is going to be. This leads to either lost sales or lot of inventory at hand. Our model's capability to provide good weekly sales forecast, evident from lower wMAPE at item-week level and figure \ref{fig:sensitivity_analysis}(a), gives an opportunity to better plan drop by moving from manual to automated drop planning driven by data and machine learning. This use case is currently being tested at our organization
\end{itemize}

\section{Conclusion and Future Directions}
We have presented the first large scale study for the demand forecast of new items in fashion. We have shown that careful feature engineering when used in conjunction with XGBoost, can be used to forecast demand for new items with reasonably good accuracies.  While creating our models and features we have been cognizant of the fact that many features will not be available as is when forecasts are being generated for future period. Hence we have used innovative transformations so that we don't have to remember train data during forecast time, thereby reducing the computation and memory requirements during forecast generation. This has also allowed our models to be easily deploy-able for internet retailers where scale and performance are crucial deciding factors in operation. 
Section~\ref{deploy} lists business results achieved corresponding to the modeling outputs realised to demonstrate real world usefulness of our work. 

Contrary to our initial expectations, DNN models (LSTM and MLP) did not show better performance over tree based models. LSTM seemed like a good choice of model theoretically since it has been shown to perform very well over various time series data, and is architecturally better suited to model long temporal dependencies. We intend to explore further in this direction by building an appropriate RNN architecture for demand forecasting that generalizes across datasets of different article types in fashion without overfitting. We are also experimenting by including image based features in our forecasting models along with currently used textual attribute embeddings. Initial results seem encouraging with image based features, but we are still working on rigorous evaluation of these models on more datasets and finding scalable ways to run such models in real world scenarios.

%
\bibliographystyle{ACM-Reference-Format}
\bibliography{sample-base}


\begin{thebibliography}{18}


\ifx \showCODEN    \undefined \def \showCODEN     #1{\unskip}     \fi
\ifx \showDOI      \undefined \def \showDOI       #1{#1}\fi
\ifx \showISBNx    \undefined \def \showISBNx     #1{\unskip}     \fi
\ifx \showISBNxiii \undefined \def \showISBNxiii  #1{\unskip}     \fi
\ifx \showISSN     \undefined \def \showISSN      #1{\unskip}     \fi
\ifx \showLCCN     \undefined \def \showLCCN      #1{\unskip}     \fi
\ifx \shownote     \undefined \def \shownote      #1{#1}          \fi
\ifx \showarticletitle \undefined \def \showarticletitle #1{#1}   \fi
\ifx \showURL      \undefined \def \showURL       {\relax}        \fi
\providecommand\bibfield[2]{#2}
\providecommand\bibinfo[2]{#2}
\providecommand\natexlab[1]{#1}
\providecommand\showeprint[2][]{arXiv:#2}

\bibitem[\protect\citeauthoryear{??}{ARI}{[n. d.]}]%
        {ARIMA_wiki}
 \bibinfo{year}{[n. d.]}\natexlab{}.
\newblock \bibinfo{title}{Autoregressive integrated moving average (ARIMA)}.
\newblock
  \bibinfo{howpublished}{\url{https://en.wikipedia.org/wiki/Autoregressive_integrated_moving_average}}.
\newblock
\newblock
\shownote{Accessed: 2019-05-02.}


\bibitem[\protect\citeauthoryear{??}{H&M}{[n. d.]}]%
        {H&M}
 \bibinfo{year}{[n. d.]}\natexlab{}.
\newblock \bibinfo{title}{H\&M, a Fashion Giant, Has a Problem: \${4.3} Billion
  in Unsold Clothes}.
\newblock
  \bibinfo{howpublished}{\url{https://www.nytimes.com/2018/03/27/business/hm-clothes-stock-sales.html}}.
\newblock
\newblock
\shownote{Accessed: 2019-05-02.}


\bibitem[\protect\citeauthoryear{??}{one}{[n. d.]}]%
        {one_hot}
 \bibinfo{year}{[n. d.]}\natexlab{}.
\newblock \bibinfo{title}{One Hot Encoding}.
\newblock
  \bibinfo{howpublished}{\url{https://scikit-learn.org/stable/modules/generated/sklearn.preprocessing.OneHotEncoder.html}}.
\newblock
\newblock
\shownote{Accessed: 2019-05-02.}


\bibitem[\protect\citeauthoryear{Bergstra, Yamins, and Cox}{Bergstra
  et~al\mbox{.}}{2013}]%
        {bergstra2013making}
\bibfield{author}{\bibinfo{person}{James Bergstra}, \bibinfo{person}{Daniel
  Yamins}, {and} \bibinfo{person}{David~Daniel Cox}.}
  \bibinfo{year}{2013}\natexlab{}.
\newblock \showarticletitle{Making a science of model search: Hyperparameter
  optimization in hundreds of dimensions for vision architectures}.
\newblock  (\bibinfo{year}{2013}).
\newblock


\bibitem[\protect\citeauthoryear{Chen and Guestrin}{Chen and Guestrin}{2016}]%
        {Chen:2016:XST:2939672.2939785}
\bibfield{author}{\bibinfo{person}{Tianqi Chen} {and} \bibinfo{person}{Carlos
  Guestrin}.} \bibinfo{year}{2016}\natexlab{}.
\newblock \showarticletitle{{XGBoost}: A Scalable Tree Boosting System}. In
  \bibinfo{booktitle}{\emph{Proceedings of the 22nd ACM SIGKDD International
  Conference on Knowledge Discovery and Data Mining}}
  \emph{(\bibinfo{series}{KDD '16})}. \bibinfo{publisher}{ACM},
  \bibinfo{address}{New York, NY, USA}, \bibinfo{pages}{785--794}.
\newblock
\showISBNx{978-1-4503-4232-2}
\urldef\tempurl%
\url{https://doi.org/10.1145/2939672.2939785}
\showDOI{\tempurl}


\bibitem[\protect\citeauthoryear{Dorogush, Ershov, and Gulin}{Dorogush
  et~al\mbox{.}}{2018}]%
        {dorogush2018catboost}
\bibfield{author}{\bibinfo{person}{Anna~Veronika Dorogush},
  \bibinfo{person}{Vasily Ershov}, {and} \bibinfo{person}{Andrey Gulin}.}
  \bibinfo{year}{2018}\natexlab{}.
\newblock \showarticletitle{CatBoost: gradient boosting with categorical
  features support}.
\newblock \bibinfo{journal}{\emph{arXiv preprint arXiv:1810.11363}}
  (\bibinfo{year}{2018}).
\newblock


\bibitem[\protect\citeauthoryear{Flunkert, Salinas, and Gasthaus}{Flunkert
  et~al\mbox{.}}{2017}]%
        {flunkert2017deepar}
\bibfield{author}{\bibinfo{person}{Valentin Flunkert}, \bibinfo{person}{David
  Salinas}, {and} \bibinfo{person}{Jan Gasthaus}.}
  \bibinfo{year}{2017}\natexlab{}.
\newblock \showarticletitle{DeepAR: Probabilistic forecasting with
  autoregressive recurrent networks}.
\newblock \bibinfo{journal}{\emph{arXiv preprint arXiv:1704.04110}}
  (\bibinfo{year}{2017}).
\newblock


\bibitem[\protect\citeauthoryear{Guo and Berkhahn}{Guo and Berkhahn}{2016}]%
        {guo2016entity}
\bibfield{author}{\bibinfo{person}{Cheng Guo} {and} \bibinfo{person}{Felix
  Berkhahn}.} \bibinfo{year}{2016}\natexlab{}.
\newblock \showarticletitle{Entity embeddings of categorical variables}.
\newblock \bibinfo{journal}{\emph{arXiv preprint arXiv:1604.06737}}
  (\bibinfo{year}{2016}).
\newblock


\bibitem[\protect\citeauthoryear{Hinton, Srivastava, Krizhevsky, Sutskever, and
  Salakhutdinov}{Hinton et~al\mbox{.}}{2012}]%
        {hinton2012improving}
\bibfield{author}{\bibinfo{person}{Geoffrey~E Hinton}, \bibinfo{person}{Nitish
  Srivastava}, \bibinfo{person}{Alex Krizhevsky}, \bibinfo{person}{Ilya
  Sutskever}, {and} \bibinfo{person}{Ruslan~R Salakhutdinov}.}
  \bibinfo{year}{2012}\natexlab{}.
\newblock \showarticletitle{Improving neural networks by preventing
  co-adaptation of feature detectors}.
\newblock \bibinfo{journal}{\emph{arXiv preprint arXiv:1207.0580}}
  (\bibinfo{year}{2012}).
\newblock


\bibitem[\protect\citeauthoryear{Ke, Meng, Finley, Wang, Chen, Ma, Ye, and
  Liu}{Ke et~al\mbox{.}}{2017}]%
        {Ke2017LightGBMAH}
\bibfield{author}{\bibinfo{person}{Guolin Ke}, \bibinfo{person}{Qi Meng},
  \bibinfo{person}{Thomas Finley}, \bibinfo{person}{Taifeng Wang},
  \bibinfo{person}{Wei Chen}, \bibinfo{person}{Weidong Ma},
  \bibinfo{person}{Qiwei Ye}, {and} \bibinfo{person}{Tie-Yan Liu}.}
  \bibinfo{year}{2017}\natexlab{}.
\newblock \showarticletitle{LightGBM: A Highly Efficient Gradient Boosting
  Decision Tree}. In \bibinfo{booktitle}{\emph{NIPS}}.
\newblock


\bibitem[\protect\citeauthoryear{Mik}{Mik}{2019}]%
        {ellen2019}
\bibfield{author}{\bibinfo{person}{Ellen~C. Mik}.}
  \bibinfo{year}{2019}\natexlab{}.
\newblock \emph{\bibinfo{title}{New Product Demand Forecasting, A Literature
  Study}}.
\newblock \bibinfo{thesistype}{Master's\ thesis}. \bibinfo{school}{Vrije
  Universitat, Amsterdam}.
\newblock
\newblock
\shownote{(In preparation).}


\bibitem[\protect\citeauthoryear{Nenni, Giustiniano, and Pirolo}{Nenni
  et~al\mbox{.}}{2013}]%
        {nenni2013demand}
\bibfield{author}{\bibinfo{person}{Maria~Elena Nenni}, \bibinfo{person}{Luca
  Giustiniano}, {and} \bibinfo{person}{Luca Pirolo}.}
  \bibinfo{year}{2013}\natexlab{}.
\newblock \showarticletitle{Demand forecasting in the fashion industry: a
  review}.
\newblock \bibinfo{journal}{\emph{International Journal of Engineering Business
  Management}}  \bibinfo{volume}{5} (\bibinfo{year}{2013}),
  \bibinfo{pages}{37}.
\newblock


\bibitem[\protect\citeauthoryear{Paszke, Gross, Chintala, Chanan, Yang, DeVito,
  Lin, Desmaison, Antiga, and Lerer}{Paszke et~al\mbox{.}}{2017}]%
        {paszke2017automatic}
\bibfield{author}{\bibinfo{person}{Adam Paszke}, \bibinfo{person}{Sam Gross},
  \bibinfo{person}{Soumith Chintala}, \bibinfo{person}{Gregory Chanan},
  \bibinfo{person}{Edward Yang}, \bibinfo{person}{Zachary DeVito},
  \bibinfo{person}{Zeming Lin}, \bibinfo{person}{Alban Desmaison},
  \bibinfo{person}{Luca Antiga}, {and} \bibinfo{person}{Adam Lerer}.}
  \bibinfo{year}{2017}\natexlab{}.
\newblock \showarticletitle{Automatic differentiation in PyTorch}. In
  \bibinfo{booktitle}{\emph{NIPS-W}}.
\newblock


\bibitem[\protect\citeauthoryear{Pedregosa, Varoquaux, Gramfort, Michel,
  Thirion, Grisel, Blondel, Prettenhofer, Weiss, Dubourg, Vanderplas, Passos,
  Cournapeau, Brucher, Perrot, and Duchesnay}{Pedregosa et~al\mbox{.}}{2011}]%
        {scikit-learn}
\bibfield{author}{\bibinfo{person}{F. Pedregosa}, \bibinfo{person}{G.
  Varoquaux}, \bibinfo{person}{A. Gramfort}, \bibinfo{person}{V. Michel},
  \bibinfo{person}{B. Thirion}, \bibinfo{person}{O. Grisel},
  \bibinfo{person}{M. Blondel}, \bibinfo{person}{P. Prettenhofer},
  \bibinfo{person}{R. Weiss}, \bibinfo{person}{V. Dubourg}, \bibinfo{person}{J.
  Vanderplas}, \bibinfo{person}{A. Passos}, \bibinfo{person}{D. Cournapeau},
  \bibinfo{person}{M. Brucher}, \bibinfo{person}{M. Perrot}, {and}
  \bibinfo{person}{E. Duchesnay}.} \bibinfo{year}{2011}\natexlab{}.
\newblock \showarticletitle{{Scikit-learn: Machine Learning in Python }}.
\newblock \bibinfo{journal}{\emph{Journal of Machine Learning Research}}
  \bibinfo{volume}{12} (\bibinfo{year}{2011}), \bibinfo{pages}{2825--2830}.
\newblock


\bibitem[\protect\citeauthoryear{Santurkar, Tsipras, Ilyas, and
  Madry}{Santurkar et~al\mbox{.}}{2018}]%
        {santurkar2018does}
\bibfield{author}{\bibinfo{person}{Shibani Santurkar},
  \bibinfo{person}{Dimitris Tsipras}, \bibinfo{person}{Andrew Ilyas}, {and}
  \bibinfo{person}{Aleksander Madry}.} \bibinfo{year}{2018}\natexlab{}.
\newblock \showarticletitle{How does batch normalization help optimization?}.
  In \bibinfo{booktitle}{\emph{Advances in Neural Information Processing
  Systems}}. \bibinfo{pages}{2483--2493}.
\newblock


\bibitem[\protect\citeauthoryear{Smith}{Smith}{2017}]%
        {smith2017cyclical}
\bibfield{author}{\bibinfo{person}{Leslie~N Smith}.}
  \bibinfo{year}{2017}\natexlab{}.
\newblock \showarticletitle{Cyclical learning rates for training neural
  networks}. In \bibinfo{booktitle}{\emph{2017 IEEE Winter Conference on
  Applications of Computer Vision (WACV)}}. IEEE, \bibinfo{pages}{464--472}.
\newblock


\bibitem[\protect\citeauthoryear{Sutskever, Vinyals, and Le}{Sutskever
  et~al\mbox{.}}{2014}]%
        {NIPS2014_5346}
\bibfield{author}{\bibinfo{person}{Ilya Sutskever}, \bibinfo{person}{Oriol
  Vinyals}, {and} \bibinfo{person}{Quoc~V Le}.}
  \bibinfo{year}{2014}\natexlab{}.
\newblock \showarticletitle{Sequence to Sequence Learning with Neural
  Networks}.
\newblock In \bibinfo{booktitle}{\emph{Advances in Neural Information
  Processing Systems 27}}, \bibfield{editor}{\bibinfo{person}{Z.~Ghahramani},
  \bibinfo{person}{M.~Welling}, \bibinfo{person}{C.~Cortes},
  \bibinfo{person}{N.~D. Lawrence}, {and} \bibinfo{person}{K.~Q. Weinberger}}
  (Eds.). \bibinfo{publisher}{Curran Associates, Inc.},
  \bibinfo{pages}{3104--3112}.
\newblock
\urldef\tempurl%
\url{http://papers.nips.cc/paper/5346-sequence-to-sequence-learning-with-neural-networks.pdf}
\showURL{%
\tempurl}


\bibitem[\protect\citeauthoryear{Thomassey and Fiordaliso}{Thomassey and
  Fiordaliso}{2006}]%
        {thomassey2006hybrid}
\bibfield{author}{\bibinfo{person}{S{\'e}bastien Thomassey} {and}
  \bibinfo{person}{Antonio Fiordaliso}.} \bibinfo{year}{2006}\natexlab{}.
\newblock \showarticletitle{A hybrid sales forecasting system based on
  clustering and decision trees}.
\newblock \bibinfo{journal}{\emph{Decision Support Systems}}
  \bibinfo{volume}{42}, \bibinfo{number}{1} (\bibinfo{year}{2006}),
  \bibinfo{pages}{408--421}.
\newblock


\end{thebibliography}

\appendix
\section{Appendix}
\setcounter{table}{0}
\renewcommand{\thetable}{A\arabic{table}}

We list down results on some more article types for different types of models/loss functions used, and find that XGBoost with an MSE loss function consistently outperforms other choice of models and loss functions.

\begin{table*}[!ht]
    \centering
    \caption{Model Performance for  Kurtas on Test Data, 9,161 time series or items}
  \label{tab:perf_kurtas_test}
    \begin{tabular}{p{3cm}| p{2cm}| p{2cm}| p{2cm}| p{2cm}| p{2cm}| p{2cm}}
    \hline
          Model & Criterion/Loss Function& \multicolumn{3}{c|}{wMAPE}&  PearsonR & Kendall Tau  \\
    \hline
        & &item-week & item & article type & &\\ 
    \hline
        Naive (Avg. style-week sales) Model& & 1.08&0.92 &0.23  &0.24 &0.36 \\
XGBoost&MSE&\textbf{0.6}&\textbf{0.46}&\textbf{0.08}&\textbf{0.85}&\textbf{0.71}\\
GBRT&Huber&0.64&0.49&\textbf{0.08}&0.83&0.69\\
LGBM&MSE&0.64&0.51&0.12&0.81&0.68\\
CatBoost&MSE&0.64&0.51&\textbf{0.08}&0.8&0.68\\
LSTM&Poisson&0.64&0.54&0.24&0.84&0.62\\
    \hline
    \end{tabular}
    
\end{table*}

\begin{table*}[!ht]
    \centering
    \caption{Model Performance for  Tops on Test Data, 10,618 time series or items}
  \label{tab:perf_tops_test}
    \begin{tabular}{p{3cm}| p{2cm}| p{2cm}| p{2cm}| p{2cm}| p{2cm}| p{2cm}}
    \hline
          Model & Criterion/Loss Function& \multicolumn{3}{c|}{wMAPE}&  PearsonR & Kendall Tau  \\
    \hline
        & &item-week & item & article type & &\\ 
    \hline
      Naive (Avg. style-week sales) Model & &1 &0.85  & 0.28&0.21 &0.39 \\
        XGBoost&MSE&\textbf{0.51}&\textbf{0.37}&0.18&0.91&\textbf{0.75}\\
GBRT&Huber&0.54&0.39&0.2&0.91&0.73\\
XGBoost&Poisson&0.54&0.38&\textbf{0.06}&\textbf{0.92}&0.71\\
CatBoost&MSE&0.54&0.4&0.2&0.91&0.71\\
GBRT&MSE&0.55&0.39&0.15&0.9&0.73\\
    \hline
    \end{tabular}
    
\end{table*}

\begin{table*}[!ht]
    \centering
    \caption{Model Performance for  Tshirts on Test Data, 18,364 time series or items}
  \label{tab:perf_tshirts_test}
    \begin{tabular}{p{3cm}| p{2cm}| p{2cm}| p{2cm}| p{2cm}| p{2cm}| p{2cm}}
    \hline
          Model & Criterion/Loss Function& \multicolumn{3}{c|}{wMAPE}&  PearsonR & Kendall Tau  \\
    \hline
        & &item-week & item & article type & &\\ 
    \hline
      Naive (Avg. style-week sales) Model &&1.01&0.86&0.27&0.3&0.4 \\
      XGBoost&MSE&\textbf{0.55}&\textbf{0.4}&0.04&\textbf{0.87}&\textbf{0.76}\\
CatBoost&MSE&0.57&0.42&\textbf{0.01}&0.84&0.72\\
LGBM&MSE&0.58&0.43&0.03&0.82&0.73\\
GBRT&Huber&0.59&0.44&0.06&0.84&0.74\\
LSTM&Poisson&0.6&0.46&0.07&0.85&0.64\\
    \hline
    \end{tabular}
    
\end{table*}

\begin{table*}[!ht]
    \centering
    \caption{Model Performance for Shirts on Train Data, 31,581 time series or items}
  \label{tab:perf_shirts_train}
    \begin{tabular}{p{3cm}| p{2cm}| p{2cm}| p{2cm}| p{2cm}| p{2cm}p{2cm}}
  \hline
          Model & Criterion/Loss Function& \multicolumn{3}{c|}{wMAPE}&  PearsonR & Kendall Tau  \\
    \hline
        & &item-week & item & article type & &\\ 
    \hline
      Naive (Avg. style-week sales) Model & &1.37 &1.01 & \textbf{0} &0.15 &0.32 \\
XGBoost&MSE&\textbf{0.3}&0.15&0.1&0.99&\textbf{0.89}\\
XGBoost&Poisson&\textbf{0.3}&\textbf{0.13}&\textbf{0}&\textbf{1}&0.87\\
GBRT&Huber&0.31&0.16&0.11&0.99&0.88\\
GBRT&MSE&0.34&0.17&0.1&0.99&0.87\\
LGBM&MSE&0.35&0.18&0.12&0.99&0.86\\
    \hline
 \end{tabular}
  
\end{table*}

\begin{table*}[!ht]
    \centering
    \caption{Model Performance for Casual Shoes on Train Data, 12,541 time series or items}
  \label{tab:perf_casual_shoes_train}
    \begin{tabular}{p{3cm}| p{2cm}| p{2cm}| p{2cm}| p{2cm}| p{2cm}| p{2cm}}
\hline
          Model & Criterion/Loss Function& \multicolumn{3}{c|}{wMAPE}&  PearsonR & Kendall Tau  \\
    \hline
        & &item-week & item & article type & &\\ 
    \hline
      Naive (Avg. style-week sales) Model & &1.33&0.98&\textbf{0}&0.21&0.27\\
GBRT&Huber&\textbf{0.27}&\textbf{0.14}&0.1&\textbf{0.99}&\textbf{0.91}\\
LGBM&Poisson&0.32&0.13&\textbf{0}&\textbf{0.99}&0.86\\
GBRT&MSE&0.33&0.16&0.1&\textbf{0.99}&0.89\\
XGBoost&MSE&0.33&0.18&0.12&0.98&0.88\\
LGBM&MSE&0.36&0.18&0.12&0.98&0.86\\
    \hline
 \end{tabular}
\end{table*}

\begin{table*}[!ht]
    \centering
    \caption{Model Performance for Kurtas on Train Data, 18,439 time series or ems}
  \label{tab:perf_kurtas_train}
    \begin{tabular}{p{3cm}| p{2cm}| p{2cm}| p{2cm}| p{2cm}| p{2cm}| p{2cm}}
\hline
          Model & Criterion/Loss Function& \multicolumn{3}{c|}{wMAPE}&  PearsonR & Kendall Tau  \\
    \hline
        & &item-week & item & article type & &\\ 
    \hline
      Naive (Avg. style-week sales) Model & &1.4&1.13&\textbf{0}&0.14&0.28\\
XGBoost&Poisson&\textbf{0.26}&\textbf{0.12}&\textbf{0}&\textbf{1}&0.86\\
GBRT&Huber&0.29&0.15&0.1&0.99&\textbf{0.88}\\
XGBoost&MSE&0.3&0.16&0.11&0.99&\textbf{0.88}\\
LGBM&Poisson&0.3&0.14&\textbf{0}&\textbf{1}&0.83\\
LSTM&Poisson&0.33&0.19&0.12&0.99&0.82\\
    \hline
 \end{tabular}
\end{table*}

\begin{table*}[!ht]
    \centering
    \caption{Model Performance for Tops on Train Data, 23,801 time series or items}
  \label{tab:perf_tops_train}
  \begin{tabular}{p{3cm}| p{2cm}| p{2cm}| p{2cm}| p{2cm}| p{2cm}| p{2cm}}
\hline
          Model & Criterion/Loss Function& \multicolumn{3}{c|}{wMAPE}&  PearsonR & Kendall Tau  \\
    \hline
        & &item-week & item & article type & &\\ 
    \hline
      Naive (Avg. style-week sales) Model & &1.37&1.04&\textbf{0}&0.14&0.32\\
XGBoost&Poisson&\textbf{0.28}&\textbf{0.12}&\textbf{0}&\textbf{1}&0.86\\
XGBoost&MSE&0.29&0.15&0.1&0.99&\textbf{0.89}\\
GBRT&Huber&0.3&0.17&0.11&0.99&0.88\\
GBRT&MSE&0.34&0.17&0.1&0.99&0.87\\
LSTM&Poisson&0.36&0.13&0.04&\textbf{1}&0.82\\
    \hline
 \end{tabular}
\end{table*}

\begin{table*}[!ht]
    \centering
    \caption{Model Performance for Tshirts on Train Data, 42,206 time series or items}
  \label{tab:perf_tshirts_train}
  \begin{tabular}{p{3cm}| p{2cm}| p{2cm}| p{2cm}| p{2cm}| p{2cm}| p{2cm}}
\hline
          Model & Criterion/Loss Function& \multicolumn{3}{c|}{wMAPE}&  PearsonR & Kendall Tau  \\
    \hline
        & &item-week & item & article type & &\\ 
    \hline
      Naive (Avg. style-week sales) Model &&1.37&0.95&\textbf{0}&0.18&0.3 \\
XGBoost&Poisson&\textbf{0.31}&\textbf{0.15}&\textbf{0}&\textbf{0.99}&0.86\\
GBRT&Huber&0.32&0.17&0.11&\textbf{0.99}&0.88\\
XGBoost&MSE&0.32&0.17&0.11&\textbf{0.99}&\textbf{0.89}\\
GBRT&MSE&0.36&0.19&0.11&0.98&0.87\\
LGBM&MSE&0.37&0.2&0.12&0.98&0.86\\
    \hline
 \end{tabular}
\end{table*}

%

\end{document}